\documentclass[iop,revtex4-1]{emulateapj}
\usepackage{graphicx}
\graphicspath{{./}{../figs/}}

\usepackage[dvipsnames,svgnames]{xcolor} 
\usepackage{natbib} 
\usepackage{hyperref}
\usepackage{amsmath}
\usepackage{multirow}
\usepackage{accents}
\hypersetup{    
  colorlinks      = true,
  linkcolor       = {black},
  linkbordercolor = {white},
  citecolor       = {black},
  citebordercolor = {white},
  urlcolor        = {blue},
  urlbordercolor  = {white},
}
\usepackage[figure,figure*]{hypcap}
\usepackage{import}

\usepackage{soul} 
\usepackage{amsmath}
\usepackage{xspace}
\usepackage{lineno}

\usepackage{soul} 
\usepackage{amsmath}
\usepackage{xspace}
\usepackage{xifthen}

\newcommand{\ie}{i.e.\xspace}
\newcommand{\eg}{e.g.\xspace}
\newcommand{\etc}{etc.\xspace}

\mathchardef\mhyphen="2D

\newcommand{\roughly}{\ensuremath{ {\sim}\,} }

\newlength{\dhatheight}

\newcommand{\code}[1]{\texttt{#1}\xspace}

\newcommand{\var}[1]{\ensuremath{#1}\xspace}

\newcommand{\unit}[1]{\ensuremath{\mathrm{\,#1}}\xspace}
\newcommand{\Gyr}{\unit{Gyr}}

\newcommand{\degree}{\ensuremath{{}^{\circ}}\xspace}

\newcommand{\asec}{\unit{arcsec}}

\newcommand{\km}{\unit{km}}
\newcommand{\pc}{\unit{pc}}
\newcommand{\kpc}{\unit{kpc}}
\newcommand{\second}{\unit{s}}

\newcommand{\Msun}{\ensuremath{M_\odot}}

\newcommand{\magn}{\unit{mag}}

\providecommand{\deg}{}
\renewcommand{\deg}{\unit{deg}}
\newcommand{\kms}{{\km\second^{-1}}}

\newcommand{\secref}[1]{Section~\ref{sec:#1}}

\newcommand{\tabref}[1]{Table~\ref{tab:#1}}

\newcommand{\figref}[1]{Figure~\ref{fig:#1}}

\newcommand{\bandvar}[2][]{%
  \ifthenelse{\isempty{#1}}{\var{#2}}{\var{#2\_#1}}%
}

\newcommand{\iauname}{{\maglites~J0644$-$5953}\xspace}
\newcommand{\picII}{{\iauname (Pic~II)}\xspace}

\newcommand{\modulus}{\ensuremath{m - M}\xspace}
\newcommand{\ra}{{\ensuremath{\alpha_{2000}}}\xspace}
\newcommand{\dec}{{\ensuremath{\delta_{2000}}}\xspace}

\newcommand{\age}{{\ensuremath{\tau}}\xspace}
\newcommand{\metal}{{\ensuremath{Z}}\xspace}

\newcommand{\ellip}{\ensuremath{\epsilon}\xspace}
\newcommand{\PA}{\ensuremath{\mathrm{P.A.}}\xspace}

\newcommand{\maglites}{{MagLiteS}\xspace}

\newcommand{\SExtractor}{\code{SExtractor}}
\newcommand{\HEALPix}{\code{HEALPix}}

\newcommand{\emcee}{\code{emcee}}

\newcommand{\TS}{\ensuremath{\mathrm{TS}}\xspace}

\begin{document} 

\title{An Ultra-Faint Galaxy Candidate Discovered in Early Data from the Magellanic Satellites Survey}

\def\andname{}

\author{
A.~Drlica-Wagner\altaffilmark{1,*},
K.~Bechtol\altaffilmark{2,3,\textdagger},
S.~Allam\altaffilmark{1},
D.~L.~Tucker\altaffilmark{1},
R.~A.~Gruendl\altaffilmark{4,5},
M.~D.~Johnson\altaffilmark{5},
A.~R.~Walker\altaffilmark{6},
D.~J.~James\altaffilmark{6},
D.~L.~Nidever\altaffilmark{7},
K.~A.~G.~Olsen\altaffilmark{7},
R.~H.~Wechsler\altaffilmark{8,9,10},
M.~R.~L.~Cioni\altaffilmark{11,12,13},
B.~C.~Conn\altaffilmark{14},
K.~Kuehn\altaffilmark{15},
T.~S.~Li\altaffilmark{1},
Y.-Y.~Mao\altaffilmark{16,17},
N.~F.~Martin\altaffilmark{18,19},
E.~Neilsen\altaffilmark{1},
N.~E.~D.~Noel\altaffilmark{20},
A.~Pieres\altaffilmark{21,22},
J.~D.~Simon\altaffilmark{23},
G.~S.~Stringfellow\altaffilmark{24},
R.~P.~van~der~Marel\altaffilmark{25},
B.~Yanny\altaffilmark{1}
}
\email{$^{*}$ kadrlica@fnal.gov}
\email{$^{\dagger}$ keith.bechtol@icecube.wisc.edu}
 
\affil{$^{1}$ Fermi National Accelerator Laboratory, P.O. Box 500, Batavia, IL 60510, USA}
\affil{$^{2}$ Wisconsin IceCube Particle Astrophysics Center (WIPAC), Madison, WI 53703, USA}
\affil{$^{3}$ Department of Physics, University of Wisconsin--Madison, Madison, WI 53706, USA}
\affil{$^{4}$ Department of Astronomy, University of Illinois, 1002 W. Green Street, Urbana, IL 61801, USA}
\affil{$^{5}$ National Center for Supercomputing Applications, 1205 West Clark St., Urbana, IL 61801, USA}
\affil{$^{6}$ Cerro Tololo Inter-American Observatory, National Optical Astronomy Observatory, Casilla 603, La Serena, Chile}
\affil{$^{7}$ National Optical Astronomy Observatory, 950 N. Cherry Ave., Tucson, AZ 85719, USA}
\affil{$^{8}$ Department of Physics, Stanford University, 382 Via Pueblo Mall, Stanford, CA 94305, USA}
\affil{$^{9}$ Kavli Institute for Particle Astrophysics \& Cosmology, P.O. Box 2450, Stanford University, Stanford, CA 94305, USA}
\affil{$^{10}$ SLAC National Accelerator Laboratory, Menlo Park, CA 94025, USA}
\affil{$^{11}$ Universit\"{a}t Potsdam, Institut f\"{u}r Physik und Astronomie, Karl-Liebknecht-Str. 24/25, D-14476 Potsdam, Germany}
\affil{$^{12}$ Leibnitz-Institut f\"{u}r Astrophysik Potsdam, An der Sternwarte 16, D-14482 Potsdam, Germany}
\affil{$^{13}$ University of Hertfordshire, Physics Astronomy and Mathematics, College Lane, Hatfield AL10 9AB, United Kingdom}
\affil{$^{14}$ Research School of Astronomy \& Astrophysics, Mount Stromlo Observatory, Cotter Road, Weston Creek, ACT 2611, Australia}
\affil{$^{15}$ Australian Astronomical Observatory, North Ryde, NSW 2113, Australia}
\affil{$^{16}$ Department of Physics and Astronomy, University of Pittsburgh, Pittsburgh, PA 15260, USA}
\affil{$^{17}$ Pittsburgh Particle Physics, Astrophysics, and Cosmology Center (PITT PACC), University of Pittsburgh, Pittsburgh, PA 15260, USA}
\affil{$^{18}$ Observatoire astronomique de Strasbourg, Universit\'e de Strasbourg, CNRS, UMR 7550, 11 rue de l'Universit\'e, F-67000 Strasbourg, France }
\affil{$^{19}$ Max-Planck-Institut f\"{u}r Astronomie, K\"{o}nigstuhl 17, D-69117 Heidelberg, Germany}
\affil{$^{20}$ Department of Physics, University of Surrey, Guildford GU2 7XH, UK}
\affil{$^{21}$ Instituto de F\'\i sica, UFRGS, Caixa Postal 15051, Porto Alegre, RS - 91501-970, Brazil}
\affil{$^{22}$ Laborat\'orio Interinstitucional de e-Astronomia - LIneA, Rua Gal. Jos\'e Cristino 77, Rio de Janeiro, RJ - 20921-400, Brazil}
\affil{$^{23}$ Carnegie Observatories, 813 Santa Barbara St., Pasadena, CA 91101, USA}
\affil{$^{24}$ Center for Astrophysics and Space Astronomy, University of Colorado, 389 UCB, Boulder, CO 80309-0389, USA}
\affil{$^{25}$ Space Telescope Science Institute, 3700 San Martin Drive, Baltimore, MD 21218, USA}

\begin{abstract}

We report a new ultra-faint stellar system found in Dark Energy Camera data from the first observing run of the Magellanic Satellites Survey (\maglites).
\iauname (Pictor II or Pic II) is a low surface brightness ($\mu = 28.5^{+1}_{-1} \magn \asec^{-2}$ within its half-light radius) resolved overdensity of old and metal-poor stars located at a heliocentric distance of $45^{+5}_{-4} \kpc$.
The physical size ($r_{1/2} = 46^{+15}_{-11} \pc$) and low luminosity ($M_V = -3.2^{+0.4}_{-0.5} \magn$) of this satellite are consistent with the locus of spectroscopically confirmed ultra-faint galaxies.
\picII is located $11.3^{+3.1}_{-0.9} \kpc$ from the Large Magellanic Cloud (LMC), and comparisons with simulation results in the literature suggest that this satellite was likely accreted with the LMC.
The close proximity of \picII to the LMC also makes it the most likely ultra-faint galaxy candidate to still be gravitationally bound to the LMC.

\end{abstract}

\maketitle

\section{Introduction}
\label{sec:intro}

The standard cosmological model generically predicts the formation of structure over a wide range of mass scales from galaxy clusters to ultra-faint galaxies. 
The Local Group offers a unique environment to search for evidence of hierarchical structure formation on the smallest scales. 

For decades authors have speculated that some of the smaller Milky Way satellites may have originated with the Large and Small Magellanic Clouds \citep[LMC, SMC; \eg,][]{Lynden-Bell:1976,D'Onghia:2008a,Sales:2011a,Nichols:2011a}.
The recent discovery of more than twenty ultra-faint ($M_V \gtrsim -8$) galaxy candidates by wide-area optical surveys including the Dark Energy Survey \citep[DES;][]{Bechtol:2015wya,Koposov:2015cua,Kim:2015c,Drlica-Wagner:2015ufc}, the Survey of the MAgellanic Stellar History \citep[SMASH;][]{martin_2015_hydra_ii}, Pan-STARRS \citep{Laevens:2015a,Laevens:2015b}, and VST ATLAS \citep{Torrealba:2016a,Torrealba:2016b} has renewed interest in identifying faint galactic companions of the Magellanic Clouds.
Indeed, 15 of the 17 candidates in the DES footprint are located in the southern half of the surveyed area, near to the Magellanic Clouds.
This inhomogeneity in the spatial distribution of satellites allows the DES data alone to exclude an isotropic spatial distribution of Milky Way satellites at the $3\sigma$-level \citep{Drlica-Wagner:2015ufc}. 
Instead, the observed distribution can be well, though not uniquely, described by an association between several of the new satellites and the Magellanic system. 
Simple models incorporating DES and SDSS observations predict that the entire sky may contain $\roughly 100$ ultra-faint galaxies with physical properties comparable to the DES satellites and that 20--30\% of these could be spatially associated with the Magellanic Clouds \citep{Drlica-Wagner:2015ufc}. 

These conclusions are largely supported by detailed simulations \citep{Deason:2015,Wheeler:2015,Yozin:2015,Jethwa:2016,Sales:2016}, which also find evidence for a Magellanic bias in the Milky Way satellite distribution.
In addition, the systemic radial velocities of several of the newly discovered satellites may be consistent with the orbit of the Clouds \citep{Koposov:2015b,Walker:2016,Jethwa:2016,Sales:2016}.

Since the Magellanic Clouds are likely on their first passage around the Milky Way \citep{Besla:2007,Busha:2011,Kallivayalil:2013}, satellite galaxies that originated with the Clouds would have formed in an environment that was rather different from the one they inhabit today. 
Comparing these systems to systems that formed around the Milky Way or far from any massive host would test environmental influences on the age, star formation history, and chemical evolution of the smallest galaxies.
Furthermore, the existence and properties of satellites of satellites can test the hierarchical structure predictions of $\Lambda$CDM.

Two low-luminosity satellites have been recently found around more isolated Local Volume analogs of the Magellanic Clouds: Antlia B around NGC 3109 \citep{Sand:2015} and MADCASH J074238+652501-dw around NGC 2403 \citep{Carlin:2016}.
Satellite-host associations are more certain in these cases relative to the Magellanic system, due to the absence of a nearby large galaxy like the Milky Way.
However, only the Magellanic Clouds are close enough to efficiently detect and characterize ultra-faint satellites.

\begin{figure*}
\center
\includegraphics[width=0.40\textwidth]{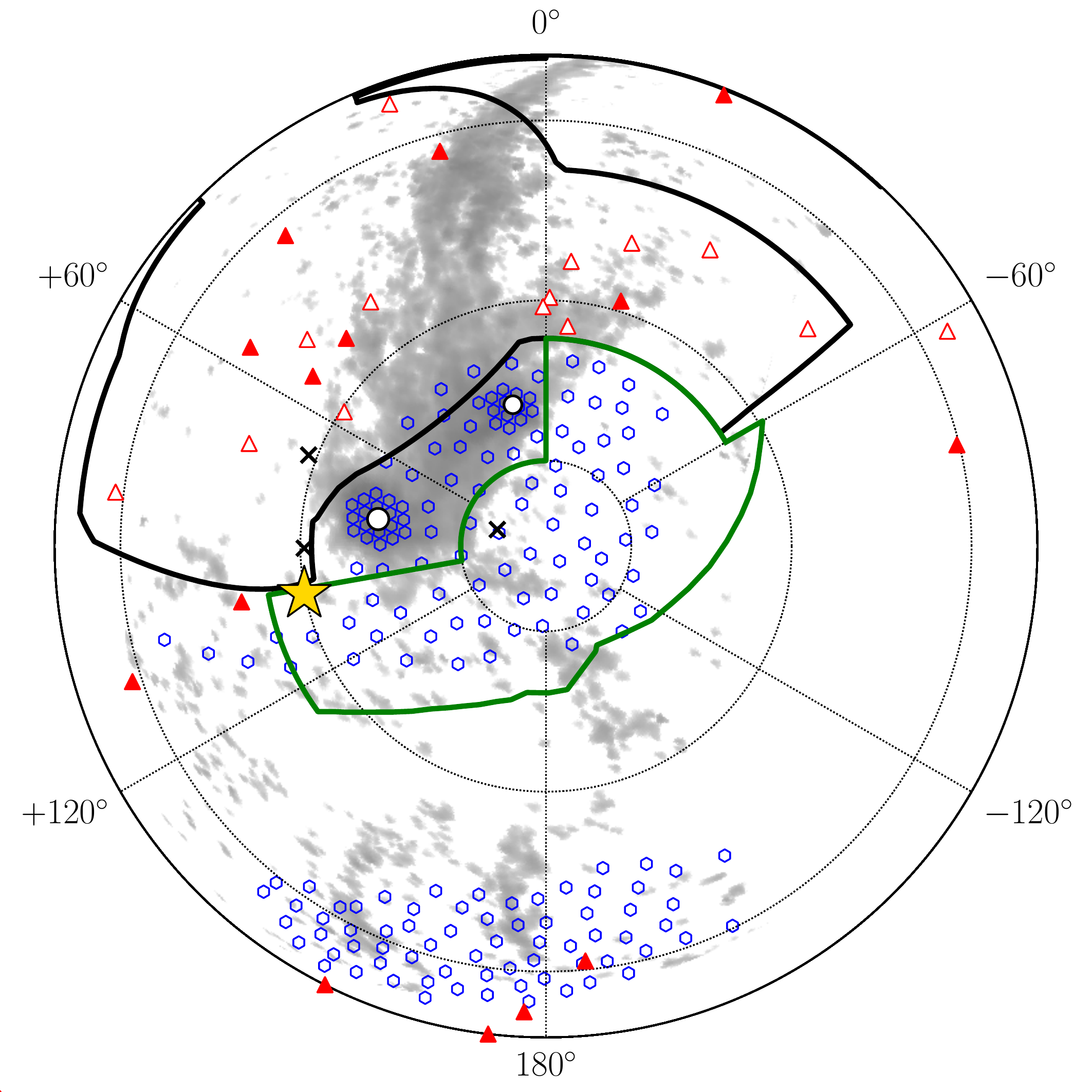}
\includegraphics[width=0.49\textwidth]{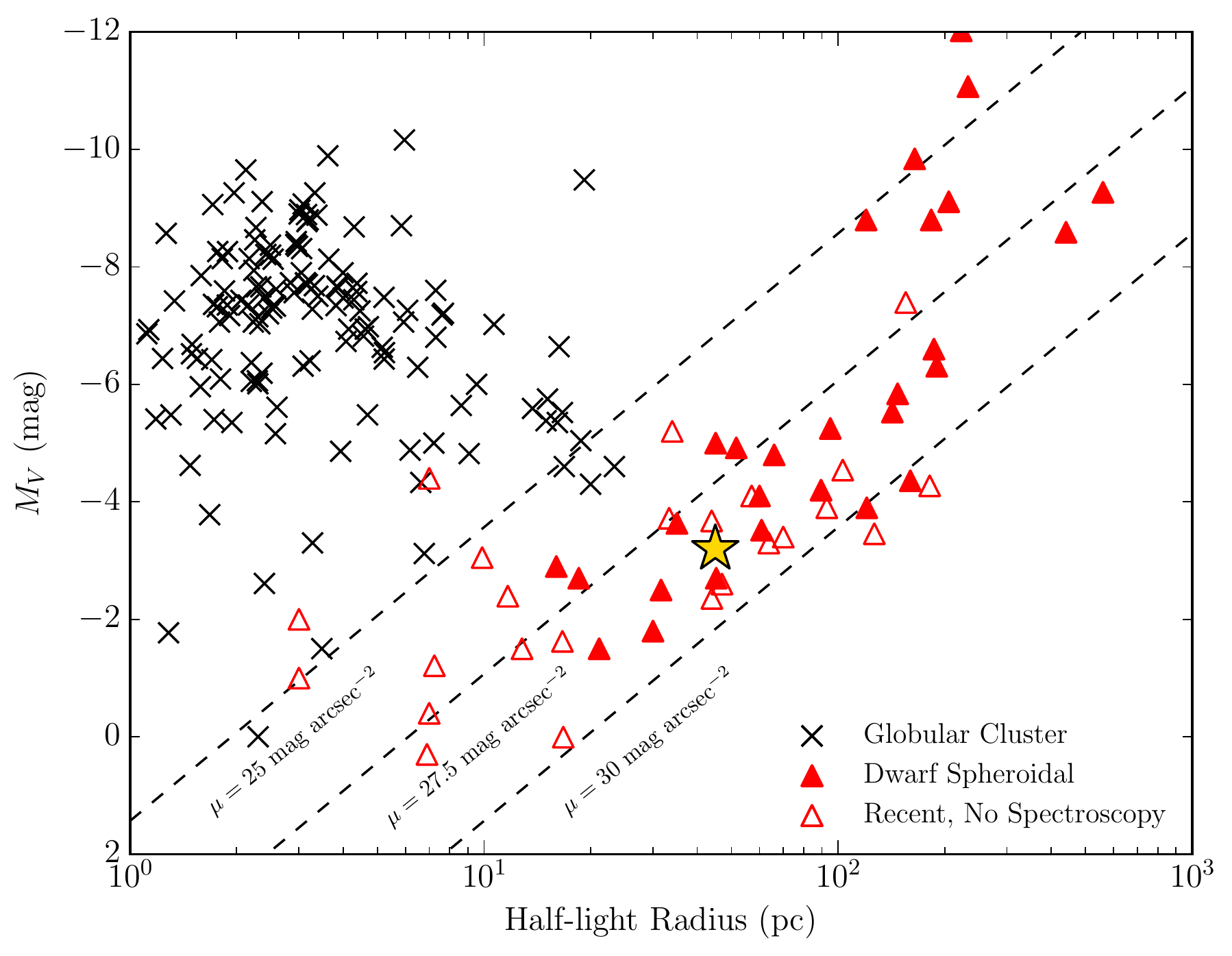}

\caption{
\textit{Left}: Orthonormal projection of the southern celestial hemisphere showing the HI density of the Magellanic Stream in gray scale \citep{Nidever:2010}. 
Over-plotted are the footprints of DES (black), \maglites (green), and SMASH (blue hexagons representing individual DECam pointings). 
The location of \picII is shown with a gold star. 
Other candidate and confirmed Milky Way satellite galaxies are marked with triangles. 
The distant LMC star clusters NGC~1841, Reticulum, and ESO~121-SC03 are marked with black crosses.
\textit{Right}:  Absolute visual magnitude ($M_V$) versus azimuthally averaged physical half-light radius ($r_{1/2}$) for dwarf galaxies (solid red triangles), globular clusters (black crosses), and recently discovered systems lacking spectroscopic measurements (open red triangles).
The black dashed lines indicate contours of constant surface brightness ($\mu$; average within the half-light radius).
\picII is marked by a gold star.
}
\label{fig:maglites}
\end{figure*}

The \textbf{Mag}ellanic Satel\textbf{Lite}s \textbf{S}urvey (\maglites; PI K. Bechtol) is a NOAO community survey that uses the Dark Energy Camera \citep[DECam;][]{Flaugher:2015} to complete an annulus of contiguous imaging around the periphery of the Magellanic system (\figref{maglites}).
In \secref{data} we describe the scope and progress of \maglites.
Initial inspection of stellar catalogs assembled from the first \maglites observing run (R1) revealed a resolved stellar overdensity at $(\ra,\dec) = (101\fdg180,-59\fdg897)$, as described in \secref{discovery}.
The physical properties of this satellite are similar to known ultra-faint galaxies (\figref{maglites}, right panel), and are detailed in \secref{properties}.
In \secref{discussion} we conclude by discussing the possible association between this stellar system and the Magellanic Clouds.

This satellite resides in the constellation Pictor, and if it is confirmed to be a dark-matter-dominated galaxy, would be named Pictor~II (Pic~II); otherwise it will be named \maglites 1. 
Until spectroscopic observations clarify the physical nature of this system, we refer to it as \picII.

\section{MagLiteS Data}
\label{sec:data}

\maglites is an ongoing optical imaging survey using DECam on the 4-m Blanco Telescope at Cerro Tololo Inter-American Observatory to map $\roughly 1200 \deg^2$ near the south celestial pole (\figref{maglites}).
\maglites relies on the large field-of-view of DECam ($\roughly 3\deg^2$) to cover this area in 12 nights distributed over the 2016A and 2017A semesters.
During this period the survey footprint will be covered with three dithered tilings.
Each tiling consists of one $90\second$ exposure in the DES $g$-band and a co-located $90\second$ exposure in the DES $r$-band such that color-magnitude diagrams can be generated.
The median $10 \sigma$ limiting depth of \maglites is $\gtrsim23 \magn$ in both bands, which is roughly comparable to the first two years of imaging by DES \citep{Drlica-Wagner:2015ufc}.

The first observing run (R1) of \maglites took place over six half-nights between 10 February 2016 and 15 February 2016.
Observing conditions were variable, with seeing between 0\farcs8 and 1\farcs5.
\maglites R1 consists of 725 survey exposures collected over an area of $\roughly 600 \deg^2$ ($\roughly 20\%$ of the exposures were in the second tiling).
Due to the southern latitude of the \maglites footprint, the airmass (and accordingly the point-spread function) of the \maglites exposures is higher than that of the DES exposures.

The \maglites exposures were reduced and processed by the DES Data Management system using the same pipeline developed for the year-three annual reprocessing of the DES data \citep[see][for an overview of the processing pipeline]{Sevilla:2011,Mohr:2012}.
Astronomical source detection and photometry were performed on a per exposure basis using the \code{PSFex} and \code{SExtractor} routines \citep{Bertin:2011, Bertin:1996}. 
As part of this step, astrometric calibration was performed with \code{SCAMP} \citep{Bertin:2006} by matching objects to the UCAC-4 catalog \citep{Zacharias:2013}. 
The \code{SExtractor} source detection threshold was set to detect sources with  $S/N \gtrsim 5$.
Photometric fluxes and magnitudes refer to the \SExtractor PSF model fit.

Unique object catalogs were assembled by performing a 1\arcsec\ match on objects detected in individual exposures.
During the catalog generation process, we flagged problematic images (\eg, CCDs suffering from reflected light, imaging artifacts, point-spread function misestimation, \etc) and excluded the affected objects from subsequent analyses.
Stellar objects were selected based on the \var{spread\_model} quantity: $|\var{wavg\_spread\_model\_r}| < 0.003 + \var{spreaderr\_model\_r}$ \citep[\eg,][]{Drlica-Wagner:2015ufc}.

Photometric calibration was performed by matching stars to the APASS catalog on a CCD-by-CCD basis.
APASS-measured magnitudes were transformed to the DES system before calibration:
\begin{align*}
g_{\rm DES} &= g_{\rm APASS} - 0.0642(g_{\rm APASS}-r_{\rm APASS}) - 0.0239 \\
r_{\rm DES} &= r_{\rm APASS} - 0.1264(r_{\rm APASS}-i_{\rm APASS}) - 0.0098
\end{align*}
For a small number of CCDs where too few stars were matched, or the resulting zeropoint was a strong outlier with respect to the other CCDs in that exposure, zeropoints were instead derived from a simultaneous fit to all CCDs on the exposure.
The zeropoints derived from APASS were found to agree well with a set of zeropoint solutions derived by the photometric standards module \citep{Tucker:2007} on four photometric nights.
Extinction from interstellar dust was calculated for each object from a bilinear interpolation of the extinction maps of \citet{Schlegel:1998}.
We followed the procedure of \citet{Schlafly:2011} to calculate reddening, assuming $R_V = 3.1$; however, in contrast to \citet{Schlafly:2011}, we used a set of $A_b/E(B-V)$ coefficients derived by DES for the $g$ and $r$ bands: $A_g/E(B-V) = 3.683$ and $A_r/E(B-V) = 2.605$.
All magnitudes quoted in the remainder of this letter have been dereddened using this procedure.

\section{Satellite Search}
\label{sec:discovery}

We performed a search for arcminute-scale stellar overdensities following the maximum-likelihood procedure described in \citet{Bechtol:2015wya}.\footnote{\url{https://github.com/DarkEnergySurvey/ugali}}
Specifically, we scanned the \maglites R1 data testing for the presence of a satellite galaxy at each location on a multi-dimensional grid of sky position (0\farcm7 resolution; \HEALPix $nside = 4096$) and distance modulus ($16 \magn <\modulus< 23 \magn$). 
Our spatial model assumed a radially symmetric Plummer kernel with half-light radius of $r_h = 4 \arcmin$.
The satellite template in color-magnitude space consisted of four \code{PARSEC} isochrones \citep{Bressan:2012} with $\age = \{10\Gyr,12\Gyr \}$ and $\metal = \{0.0001,0.0002\}$ each weighted by a \citet{Chabrier:2001} initial mass function.

\begin{figure*}
\center
\includegraphics[width=\textwidth]{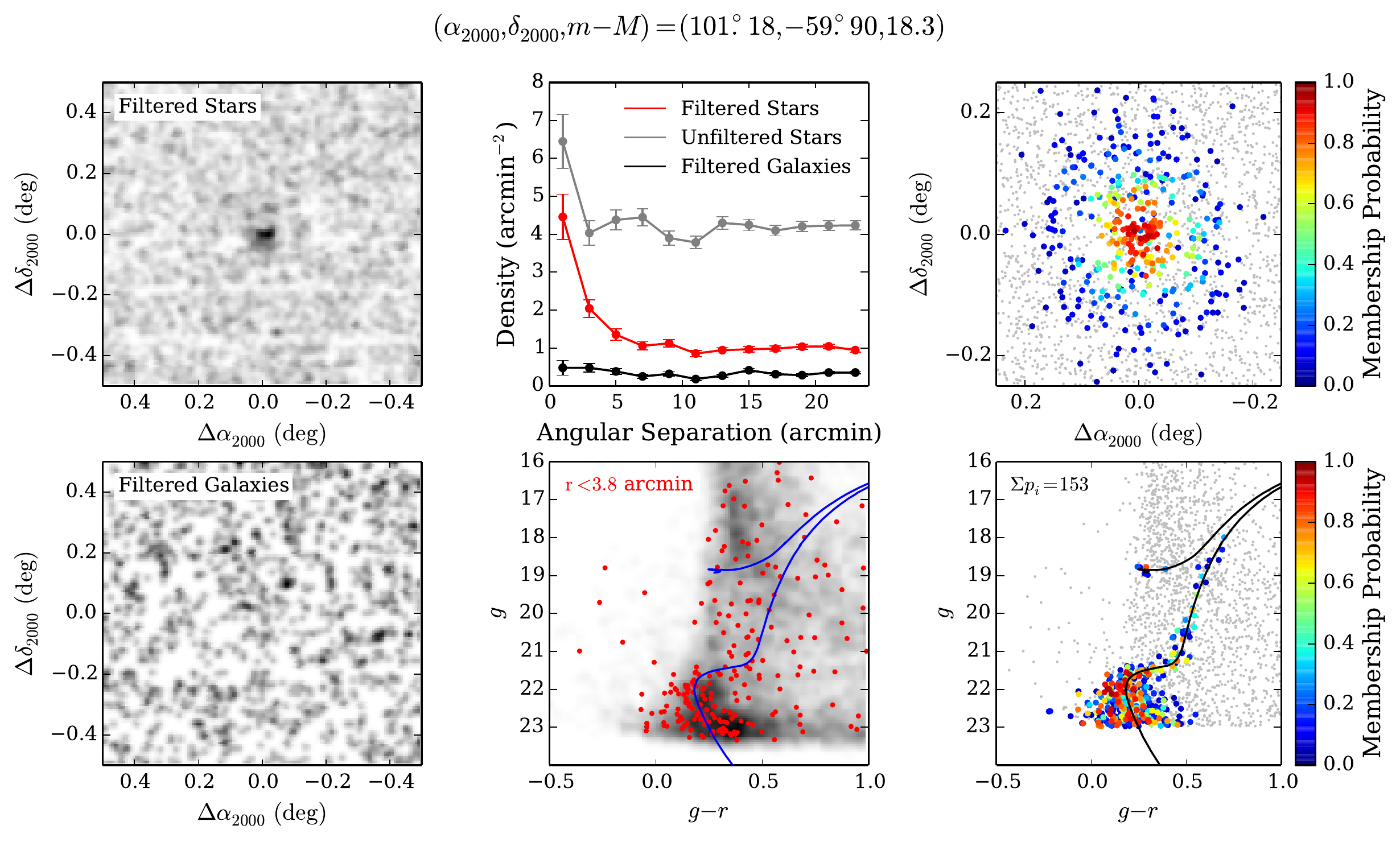}
\caption{Stellar density and color-magnitude diagrams for \picII. 
  \textit{Top left}: Spatial distribution of stars with $g < 24 \magn$ that pass the isochrone filter (see text). The field of view is $1\degree \times 1\degree$ centered on the candidate and the stellar distribution has been smoothed by a Gaussian kernel with standard deviation $0\farcm6$.
  \textit{Top center}: Radial distribution of objects with $g - r < 1 \magn$ and $g < 24 \magn$: stars passing the isochrone filter (red), stars excluded from the isochrone filter (gray), and galaxies passing the isochrone filter (black).
  \textit{Top right}: Spatial distribution of stars with high membership probabilities within a $0\fdg5 \times 0\fdg5$ field of view. Gray points indicate stars with a membership probability less than 5\%.
   \textit{Bottom left}: Same as top left panel, but for galaxies passing the isochrone filter.
   \textit{Bottom center}: The color-magnitude distribution of stars within $r = 3\farcm8$ of the centroid are indicated with individual points.
The density of the field within an annulus $0\fdg5 < r < 1\fdg0$ is represented by the background grayscale. 
The blue curve shows the best-fit isochrone as described in \secref{properties} and \tabref{properties}. 
   \textit{Bottom right}: Color-magnitude distribution of high membership probability stars.
}
\label{fig:6panel}
\end{figure*}

As noted in \secref{data}, the \maglites R1 data predominantly consists of a single DECam tiling. 
This leads to a complicated coverage pattern including gaps between CCDs, regions of significantly decreased depth due to cloudy conditions, and holes caused by scattered light from bright stars.
The maximum-likelihood analysis is capable of accounting for inhomogeneities in survey coverage, as long as they are well-characterized by a survey coverage mask.
However, inconsistencies between the parameterized coverage mask and the true survey coverage can lead to spurious detections.
While these artifacts are easily identified from visual inspection of their morphological and photometric properties, their prevalence made a systematic search of the early \maglites data difficult.
We expect these issues to be mitigated by increased survey uniformity from upcoming observations. 
Here, we focus on the fortuitous discovery of a previously unidentified stellar overdensity, \picII, which was identified by the likelihood search and passed all visual inspection tests (\figref{6panel}).
This system was identified with a likelihood ratio test statistic $\TS =  235$, which corresponds to a Gaussian significance of $\roughly 15\sigma$.

\section{Properties of \picII}
\label{sec:properties}

We simultaneously fit the morphological and isochrone parameters of \picII with the same maximum-likelihood approach used for our search.
We performed a nine-parameter fit of stellar richnesss ($\lambda$), centroid position (\ra,\dec), semi-major half-light radius ($a_h$), ellipticity (\ellip), position angle (\PA), distance modulus (\modulus), age (\age), and metallicity (\metal) of the stellar population.
We explored this multi-dimensional parameter space with $\roughly 8 \times 10^5$ samples from an affine invariant Markov Chain Monte Carlo (MCMC) ensemble sampler \citep{Foreman-Mackey:2013}.\footnote{\emcee v2.2.0: \url{http://dan.iel.fm/emcee/}}
We imposed a Jeffreys' prior on the extension, $\mathcal{P}(a_h) \propto 1/a_h$, and flat priors on all other parameters.
During the fit, we used a single \code{PARSEC} isochrone with age and metallicity varying between $1 \Gyr < \age < 13.5 \Gyr$ and $0.0001 < \metal < 0.01$, respectively.
Masking a bright star $4\farcm4$ southeast of \picII may bias the fit at the 1\% level.

\begin{figure*}
\center
\includegraphics[width=\textwidth]{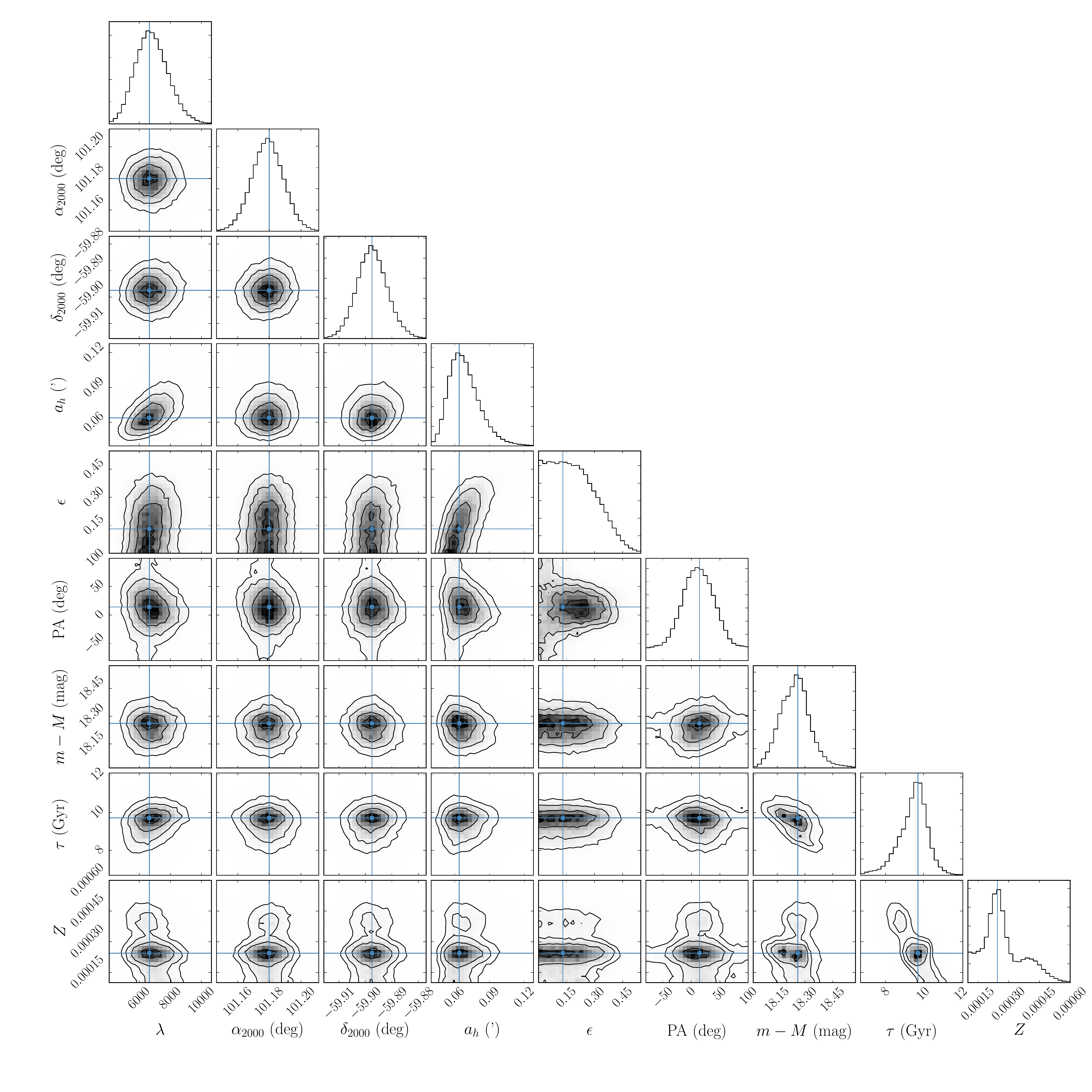}
\caption{
Posterior probability densities from a nine-parameter maximum-likelihood fit of \picII. 
From left to right the nine parameters are: stellar richness ($\lambda$), right ascension (\ra), declination (\dec), semi-major half-light radius ($a_h$), ellipticity (\ellip), position angle (\PA), distance modulus (\modulus), age (\age), and metallicity (\metal).
The crosshairs indicate the best-fit parameter values from a kernel density estimate of the peak of the posterior distribution.
}
\label{fig:mcmc}
\end{figure*}

The resulting posterior probability distributions are shown in \figref{mcmc}.
Best-fit parameters were derived from the peak of the posterior probability distribution as determined by a kernel density estimate, while uncertainties were derived from the maximum density interval that encloses 90\% of the posterior density.
The absolute $V$-band magnitude was calculated according to the prescription of \citet{Martin:2008} and does not include the uncertainty on distance.
The properties of \picII are collected in \tabref{properties}.

Like the other parameters shown in \tabref{properties}, the distance modulus of \picII was derived from a simultaneous likelihood fit to the CMD and spatial information.
The best-fit distance modulus was driven by the position of the main sequence turn-off and was only moderately influenced by potential member stars in the horizontal branch.
The statistical uncertainty on the distance modulus of \picII was derived from the posterior probability distribution, marginalizing over the other parameters (most importantly the age and metallicity).
There is an additional systematic uncertainty coming from the synthetic isochrone model. 
Fitting the data with synthetic isochrones from \citet{Dotter:2008} decreased the distance modulus by $0.05 \magn$.
This variation is consistent with the results of previous work \citep{Koposov:2015cua, Drlica-Wagner:2015ufc}, and we quote a systematic uncertainty on the distance modulus of $\pm 0.1 \magn$ associated with the isochrone model.

\figref{6panel} shows the spatial and color-magnitude distribution of stellar objects surrounding \picII.
To enhance contrast with the field population, we filter in color-magnitude space by selecting catalog objects within $0.1\magn$ of the best-fit old and metal-poor \code{PARSEC} isochrone ($\age = 10\Gyr$, $\metal = 0.0002$).
The rightmost panels show the spatial and color-magnitude distributions of stars in the region color-coded with the membership probability assigned by the likelihood analysis.
In addition to a densely populated main sequence, \picII has a handful of probable members on the red giant branch and a few possible horizontal branch members.

\begin{deluxetable}{c c c}
\tablewidth{0pt}
\tabletypesize{\small}
\tablecaption{
\label{tab:properties}
Observed and derived properties of \picII \label{tab:picII}
}
\tablehead{
Parameter & Value & Units
}
\startdata
\ra,\dec & 101.180,-59.897 & deg,deg \\
$a_h$ & $3.8^{+1.5}_{-1.0}$ & arcmin \\
$r_h$ & $3.6^{+1.5}_{-0.9}$ & arcmin \\
\ellip & $0.13^{+0.22}_{-0.13}$ & \ldots \\
\PA & $14^{+60}_{-66}$ & deg \\
\modulus & $18.3^{+0.12}_{-0.15} \pm 0.1$\tablenotemark{a} & \ldots \\
\age & $10^{+1}_{-2}$  & \Gyr \\
\metal & $0.0002^{+0.0003}_{-0.0001}$ & \ldots \\
$\sum p_i$ & $153^{+12}_{-12}$ & \ldots \\
\TS & $235$ & \ldots \\

\multicolumn{3}{c}{ \hrulefill } \\

$D_{\odot}$ & $45^{+5}_{-4}$ & kpc \\
$r_{1/2}$ & $46^{+15}_{-11}$ & pc \\
$M_V$ & $-3.2^{+0.4}_{-0.5}$\tablenotemark{b} & mag \\
$M_{*}$ & $1.6^{+0.5}_{-0.3}$ & $10^3 \Msun$ \\
$\mu$ & $28.5^{+1}_{-1}$ & mag/arcsec${}^2$ \\
${\rm [Fe/H]}$ & $-1.8^{+0.3}_{-0.3}$ & dex \\
$E(B-V)$ & 0.107 & mag \\

\multicolumn{3}{c}{ \hrulefill } \\

$D_{\rm LMC}$ & $11.3^{+3.1}_{-0.9}$ & kpc \\
$D_{\rm SMC}$ & $35^{+3}_{-2}$ & kpc \\
$D_{\rm GC}$  & $45^{+5}_{-4}$ & kpc \\

\enddata
\tablecomments{Uncertainties were derived from the highest density interval containing the peak and 90\% of the marginalized posterior distribution.}
\tablenotetext{a}{We assume a systematic uncertainty of $\pm0.1$ associated with isochrone modeling.}
\tablenotetext{b}{The uncertainty in $M_V$ was calculated following \citet{Martin:2008} and does not include uncertainty in the distance.}
\end{deluxetable}

\section{Discussion}
\label{sec:discussion}

The low luminosity ($M_V = -3.2$) and large physical size ($r_{1/2} = 46 \pc$) of \picII are consistent with the population of dark-matter-dominated Milky Way satellite galaxies (\figref{maglites}). 
Specifically, \picII possesses structural properties similar to the recently confirmed dwarf galaxies Reticulum~II and Horologium~I \citep{Bechtol:2015wya,Koposov:2015cua}.
While stellar kinematic data are necessary to measure the dark matter content of \picII and assign a definitive classification, the \maglites photometry suggests that it will likely join the ranks of recently discovered dwarf galaxies.

The proximity between \picII and the LMC, $D_{\rm LMC} = 11.3^{+3.1}_{-0.9} \kpc$, is suggestive of a physical association between these two systems.
Several studies have shown that the population of old LMC stars extends to radii $>13\kpc$ \citep{Munoz:2006,Majewski:2009,Saha:2010,Balbinot:2015,Mackey:2016}.
Additionally, kinematic measurements by \citet{vanderMarel:2014} suggest that the LMC tidal radius is at least $16 \kpc$ and may be as large as $22 \pm 5 \kpc$, which places \picII well within the LMC sphere of influence.
The most distant LMC star clusters reside at similar distances: NGC 1841 at $D_{\rm LMC} = 14.9 \kpc$, Reticulum at $D_{\rm LMC} = 11.4 \kpc$, and ESO 121-SC03 $D_{\rm LMC} = 9.7 \kpc$ \citep{Schommer:1992}.
If \picII is bound to the LMC it would be expected to have a line-of-sight velocity that is similar to these clusters: $214\kms$, $243\kms$, and $309\kms$, respectively \citep[][and references therein]{Schommer:1992}.
Incidentally, \picII is located at a heliocentric distance that is consistent with the plane of the LMC disk \citep[$\roughly 46 \kpc$;][]{vanderMarel:2014}.

Several recent studies have used numerical simulations to investigate the evolution of the Magellanic system as it was accreted onto the Milky Way \citep[\ie,][]{Deason:2015,Jethwa:2016}.
Using the ELVIS simulations \citep{Garrison-Kimmel:2014}, \citet{Deason:2015} find that $> 40\%$ of satellites galaxies that are currently located at $D_{\rm LMC} < 20 \kpc$ were bound to the LMC before infall into the Milky Way.
This fraction increases to $> 65\%$ if the Magellanic group was accreted recently ($\age_{\rm infall}  < 2 \Gyr$) and $> 80\%$ when considering only dynamical analogs of the LMC.
Based on these results, if \picII originated as a member of the LMC group then it would have a radial velocity that is within $\roughly 150 \kms$ of that of the LMC.

\begin{figure}
\includegraphics[width=\columnwidth]{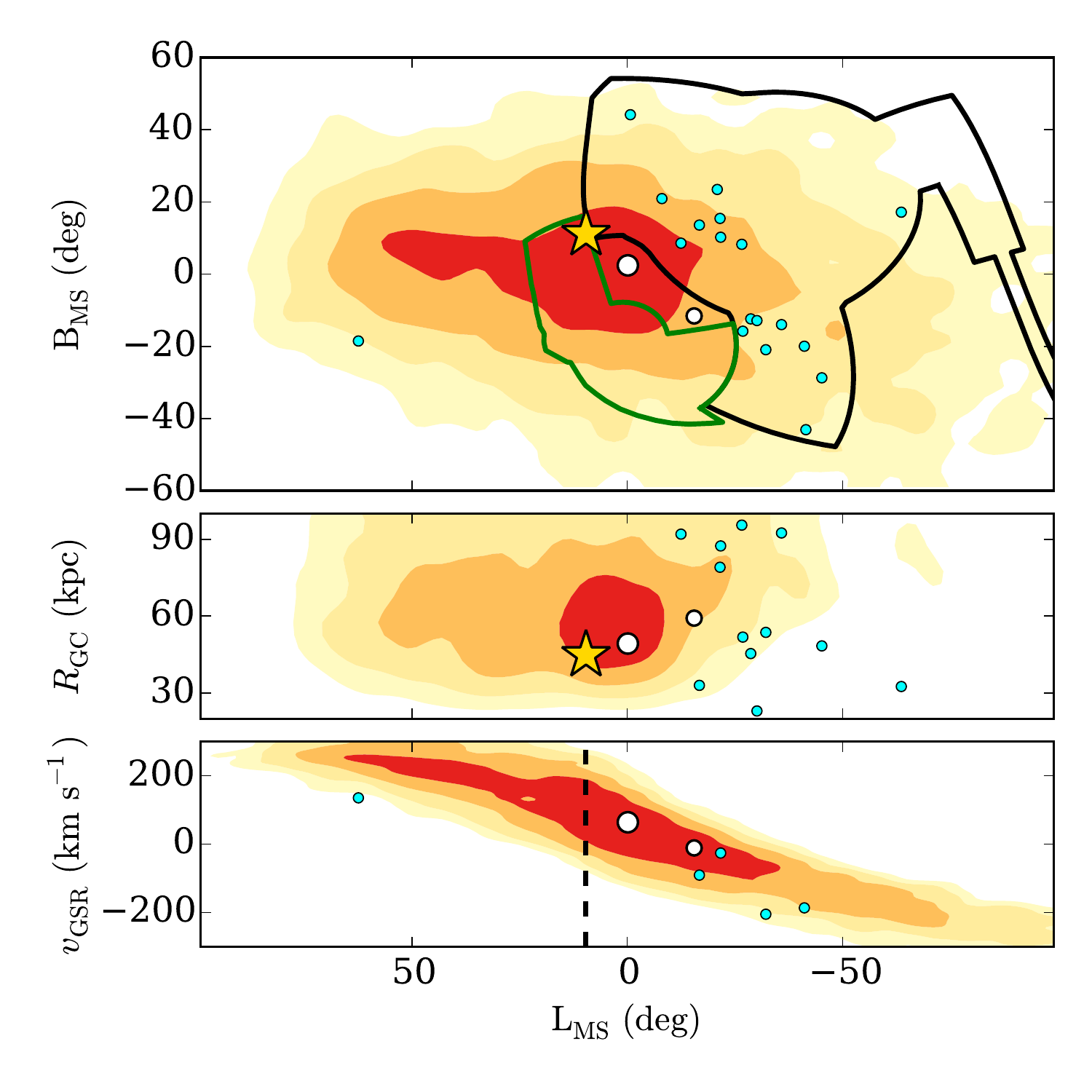}
\caption{
\label{fig:sims}
Phase space coordinates of \picII (gold star) relative to the simulated distribution of LMC satellites from \citet{Jethwa:2016}, represented by colored contours.
Recently discovered DES satellites and Hydra II are shown with cyan markers.
\textit{Top}: The density of simulated LMC satellites projected onto the sky in MS coordinates. 
The DES and \maglites footprints are outlined in black and green respectively.
\textit{Middle}: The density of simulated LMC satellites with $5\degree < {\rm B_{MS}} < 25\degree$ projected onto the plane of Galactocentric radius and MS longitude.
\textit{Bottom}: Distribution of line-of-sight velocities in the Galactocentric standard reference frame for simulated satellites of the LMC. The black dashed line represents the MS longitude of \picII.
Figure adapted from \citet{Jethwa:2016}.
}
\label{fig:association}
\end{figure}

\citet{Jethwa:2016} used dedicated simulations to model the dynamics of the Milky Way, LMC, and SMC, and concluded that $30\%$ of the Milky Way's satellite galaxies originated with the LMC.
They predict that satellites of the LMC are distributed within $\pm 20\degree$ of the plane of the Magellanic Stream \citep[MS;][]{Nidever:2008} and would be preferentially found at positive MS longitudes in a leading arm of satellites (\figref{sims}).
The MS coordinates of \picII, $(L_{\rm MS}, B_{\rm MS}) = 9\fdg58, 11\fdg11$, lie within the preferred region for Magellanic satellites and are well-aligned with the putative plane connecting the LMC, SMC, and the DES-discovered satellites with $B_{\rm MS} < 0\degree$ \citep{Jethwa:2016}.
Furthermore, the simulations of \citet{Jethwa:2016} predict that \picII has a line-of-sight velocity in the Galactic standard of rest (GSR) in the range of $15 \kms < v_{\rm GSR} < 175 \kms$ (68\% interval).

Taken together, the photometric properties of \picII and recent simulations of the Magellanic system support the hypothesis that \picII is a dwarf galaxy that arrived at the Milky Way as part of the Magellanic system.
However, kinematic measurements are required to confirm the past or present relationship between \picII and its massive nearby neighbors.
If \picII is confirmed to be a gravitationally bound galactic companion of the LMC, it would be the most direct example of a satellite of a satellite within the Local Group, further supporting the standard cosmological framework of hierarchical structure formation.
The fortuitous discovery of \picII in early \maglites data will be followed by more comprehensive searches for satellite galaxies once additional data are collected.

\section{Acknowledgments}

We thank Prashin Jethwa for sharing results from his simulated model of LMC satellites.

This project used data obtained with the Dark Energy Camera (DECam), which was constructed by the Dark Energy Survey (DES) collaboration. 
We gratefully acknowledge the DES data management group at NCSA for processing the images. 
The DES data management system is supported by the National Science Foundation under Grant Number AST-1138766.

Funding for the DES Projects has been provided by the DOE and NSF (USA), MISE (Spain), STFC (UK), HEFCE (UK), NCSA (UIUC), KICP (U. Chicago), CCAPP (Ohio State), MIFPA (Texas A\&M), CNPQ, FAPERJ, FINEP (Brazil), MINECO (Spain), DFG (Germany) and the collaborating institutions in the Dark Energy Survey, which are Argonne Lab, UC Santa Cruz, University of Cambridge, CIEMAT-Madrid, University of Chicago, University College London, DES-Brazil Consortium, University of Edinburgh, ETH Z{\"u}rich, Fermilab, University of Illinois, ICE (IEEC-CSIC), IFAE Barcelona, Lawrence Berkeley Lab, LMU M{\"u}nchen and the associated Excellence Cluster Universe, University of Michigan, NOAO, University of Nottingham, Ohio State University, University of Pennsylvania, University of Portsmouth, SLAC National Lab, Stanford University, University of Sussex, and Texas A\&M University.

This research was made possible through the use of the AAVSO Photometric All-Sky Survey (APASS), funded by the Robert Martin Ayers Sciences Fund.

{\it Facilities:} \facility{Blanco}

\end{document}